# Shift Current Anomalous Photovoltaics in a Double Perovskite Ferroelectric

Linjie Wei[a,c†], Fu Li[b†], Yi Liu[a,c*], Hongbin Zhang[b*], Junhua Luo[a,c], and Zhihua Sun[a,c,d*]

[a] State Key Laboratory of Functional Crystals and Devices, Fujian Institute of Research on the Structure of Matter, Chinese Academy of Sciences, Fuzhou, Fujian 350002, People's Republic of China.

[b] Department of Materials and Earth Sciences, Technical University of Darmstadt, Darmstadt, Germany.

[c] University of Chinese Academy of Sciences, Chinese Academy of Sciences, Beijing, 100039, People's Republic of China.

[d] Fujian Science & Technology Innovation Laboratory for Optoelectronic Information of China, Fuzhou, Fujian 350108, People's Republic of China.

**Email:** sunzhihua@fjirsm.ac.cn

[†]These authors contributed equally to this work.

**Author Contributions:** H. B. Z., J. H. L. and Z. H. S. conceived the idea. L. J. W., Y. L. and F.L. conducted the experiment. L. J. W., Y. L. and F. L. analyzed the data together with all other authors. L. J. W., Y. L., F. L. and Z. H. S. wrote the paper with contributions from all the other authors.

**Competing Interest Statement:** The authors declare no competing interests.

**Abstract**

Ferroelectric anomalous photovoltaic (APV) effect, as a fascinating physical conceptual phenomenon, holds significant potentials for new optoelectronic device applications. However, due to the knowledge lacking on the origin and underlying mechanism of ferroelectric APV effect, substantial challenges still remain in exploring new APV-active candidate materials. The emerging shift current model, involving the transfer of photogenerated charges through the displacement of wave functions, has attracted considerable attention for its unique insights into the bulk photovoltaic effect. Here, we present strong APV properties in a high-temperature double perovskite ferroelectric (cyclohexylmethylammonium)$_2$CsAgBiBr$_7$, showing an extremely large above-bandgap photovoltage up to ~40 V. This figure-of-merit is far beyond its bandgap of ~2.3 eV and comparable to the state-of-art molecular ferroelectrics. Strikingly, the shift current model reveals an intrinsic correlation with Cs$^+$ cation displacement and provides, for the first time, an explicit explanation for the structural origin of ferroelectric APV activities. Besides, its steady-state APV photocurrent exhibits the unique light-polarization dependence, which endows remarkable polarization-sensitivity with the highest polarization ratios of ~41 among the known 2D single-phase materials. As the unprecedented exploration of ferroelectric APV characteristics illuminated by the shift current mechanism, this finding paves a pathway to assemble new optoelectronic smart devices.

**Significance Statement**

Ferroelectric anomalous photovoltaic (APV) effect has holds a great promise in the new optoelectronic application fields. Nevertheless, the limited understanding of ferroelectric APV effect mechanism hinders the exploration of novel APV-active candidates. Recently, the shift current model exhibits unique insights into the bulk photovoltaic effect and has attracted considerable attention. In this work, we present strong APV properties in a high-temperature double perovskite ferroelectric, showing an extremely large above-bandgap photovoltage of ~40 V ($E_g$ ~2.3 eV), which is comparable to the state-of-art molecular ferroelectrics. Strikingly, the shift current model explicitly reveals the structural origin of ferroelectric APV activities and demonstrates the intrinsic correlation with Cs$^+$ cation displacement. This pioneering work paves a new pathway to develop new optoelectronic smart devices.

**Introduction**

Bulk photovoltaic effect refers to the physical phenomenon in which the non-centrosymmetric crystalline materials generate a steady open-circuited voltage ($V_{oc}$) and short-circuited current ($I_{sc}$) under intrinsic light absorption, holding significant application prospects in the field of sustainable optoelectronics [1, 2]. Especially, ferroelectric anomalous photovoltaic (APV) effect can induce extremely large open-circuit photovoltage ($V_{oc}$) far beyond its band gap, thus paving a potential way to realize high-performed photodetection [3]. The improvement of power conversion efficiency can also be expected by virtue of ferroelectric APV effect. Despite great potentials, the exploration of this fascinating physical phenomenon is almost limited to few insulating ferroelectric oxides, such as LiNbO$_3$ [4], BaTiO$_3$ [5] and BiFeO$_3$ [6]. These materials possess the common characteristics that their ferroelectricity originates from asymmetric spontaneous ionic displacements within the crystal lattice, which induce symmetry breaking and generate macroscopic spontaneous polarization ($P_s$). Studies on the new APV-active ferroelectric candidates remain scarce, due to the knowledge lacking on the complicated physical mechanism of APV properties. Within this portfolio, different mechanisms have been proposed, including depolarization electric field effects, optical-rectification effects and asymmetry potentials; these concepts initially presumed the built-in electrostatic field induced by spontaneous electric polarization as the origin of APV effect [7]. However, the classical theories are difficult to provide consistent explanations for the certain physical characteristics, such as the direction of photocurrent associated with the light-polarization and photon energy [8].

Currently, shift current model has emerged as a new theoretical mechanism that affords an appropriate illumination of bulk photovoltaics. The emergence of shift current arises from the phase shift of photoexcited electron wavefunction in non-centrosymmetric materials, which leads to spatial charge displacement and direct current [9]. Its fundamental principle stems from the asymmetry of materials and the Berry curvature effect, representing a significant nonlinear optical process [10]. Therefore, shift current intrinsically involves with electric polarization and is further influenced by the displacement of charged atoms or molecules. It is particularly noteworthy that asymmetric ionic displacements in the crystal lattice can significantly modulate the phase evolution of electron wavefunctions, thereby affecting the generation efficiency of shift current. Owing to these unique properties, shift current manifests as a topological current with appealing characteristics, such as ultrafast response, high quantum efficiency, low noise, and bias-free operation; these attributes make it highly advantageous for applications in photodetection [11]. Despite its numerous advantages, exploration of APV-active ferroelectric candidates with inspired by shift current model remains a significant scientific challenge. For instance, 2D perovskite ferroelectrics hold promise as an ideal platform for investigating the APV behaviors, due to their unique structural flexibility and excellent physical properties, e.g., high absorption coefficient, long carrier mobility and low trap density [12]. However, as we know, the shift current model for APV effect in 2D perovskite ferroelectrics has not been explored in detail.

Motivated by the shift current theory of bulk APV effect, a hint to achieve new APV-active ferroelectrics is to introduce ionic displacement, which could induce symmetry breaking for ferroelectricity and directly generate shift current through modulation of electronic wavefunction phases [13]. Based on this concept, the double perovskite motif of {$Cs_4AgBiBr_8$} may be an ideal platform due to the motion of $Cs^+$ cations under external stimulation. Here, we explore strong APV effect in a double perovskite ferroelectric $(CHMA)_2CsAgBiBr_7$ (**1**, where $CHMA^+$ is cyclohexylmethylammonium), which enables an extremely large photovoltage (~40 V) much higher than its band gap ($E_g$ ~2.3 eV). To our best knowledge, this is the first demonstration of such intriguing photovoltaics in 2D double perovskite ferroelectrics. Further, it is disclosed that the APV effect of **1** involves with the shift current mechanism, which is closely related to the ion migration induced by $Cs^+$ cation displacement. This ferroelectric APV activity gives rise to intriguing polarization-sensitive photodetection with high ratios up to ~41. These findings highlight the potentials of double perovskites as APV-active ferroelectric candidates toward high-performance polarized-light detection.

**Results**

**Crystal structure and ferroelectric property.** Large single crystals of compound **1** were successfully grown from concentrated aqueous HBr solution using a controlled temperature cooling approach (Fig. S1). Structural analysis reveals that **1** adopts a typical 2D Ruddlesden-Popper motif and crystallizes in the polar space group *Ama*2 at room temperature (14). As shown in Fig. 1a, the distorted $AgBr_6$ and $BiBr_6$ octahedra alternate to form the inorganic frameworks. The ordered $CHMA^+$ cations serve as the organic spacers and connect to the inorganic sheets through strong N-H···Br hydrogen bonds, aligning along the crystallographic *c*-axis direction. Interestingly, the $Cs^+$ cations are located within the internal cavities of inorganic framework, which also displace from the central positions of cavities along the *c*-axis direction. According to the calculated average atomic coordinates within the unit cell, the $Cs^+$ cations are located at (0.5, 0.5, 0.644) with a large displacement ($\Delta d$) of ~0.144 Å, compared to the central position (0.5, 0.5, 0.5) of the paraelectric phase structure (Figs. 1b and S2). This displacement of $Cs^+$ cations is expected to induce electric polarization and even ferroelectricity in **1**. Fig. S3 shows that the symmetry breaking of **1** conforms to the Aizu rule of 4/*mmmFmm*2, revealing its biaxial characteristics of ferroelectricity. To further elucidate the ferroelectricity of **1**, we measured polarization *versus* electric field (*P-E*) hysteresis loops and current *versus* electric field (*J-E*) curves at different temperatures, as shown in Fig. 1c and 1d. The results show that its $P_s$ and coercive electric field ($E_c$) at 300 K are approximately ~4.3 $\mu C\ cm^{-2}$ and ~88 $kV\ cm^{-1}$, which are comparable to recently reported 2D perovskite ferroelectrics, such as (*n*-butylammonium)$_2$(methylammonium)$Pb_2Br_7$ ($P_s$ = 3.6 $\mu C\ cm^{-2}$) (15), (isoamylammonium)$_2$(ethylammonium)$_2Pb_3I_{10}$ ($P_s$ = 5.2 $\mu C\ cm^{-2}$) (16) and (*n*-butylamine)$_2$(ethylamine)$_2Pb_3I_{10}$ ($P_s$ = 4.1 $\mu C\ cm^{-2}$) (17). Moreover, the $P_s$ and $E_c$ gradually decrease with the increasing of temperature until 338 K (Fig. S4). It is noteworthy that **1** has excellent fatigue durability under electric field, and its $P_s$ value remains stable after 4 × $10^5$ switching operation cycles (Fig. 1e). Since that the built-in electrostatic field generated by ferroelectricity is beneficial to the separation and migration of charge carriers, the displacement behavior of $Cs^+$ cations within the inorganic cavities potentially offers an opportunity to trigger bulk photovoltaic effect.

**Ferroelectric APV effect.** Compound **1** exhibits a sharp absorption onset at 525 nm, corresponding to an optical bandgap ($E_g$) of approximately 2.3 eV, in agreement with theoretical predictions (Fig. S5). Subsequently, the single crystal with dimension of ~2 × 2 × 0.5 $mm^3$ was chosen to fabricate a lateral photovoltaic device. As shown in Fig. 2a, two orthogonal sets of silver electrodes were thermally evaporated onto the (100) crystal surface, to measure the bulk photovoltaic effects parallel to the *c*- and *b*-axes (expressed as $E_{\parallel c}$ and $E_{\parallel b}$), respectively. As shown in the voltage-current (*I-V*) curves under illumination (405 nm, 149 $mW/cm^2$), **1** exhibits a large output photovoltage ($V_{oc}$) of ~40 and 38 V in the direction of $E_{\parallel c}$ and $E_{\parallel b}$; the corresponding short-circuit current densities ($J_{sc}$) are approximately 15 and 13 $nA/cm^2$, respectively (Fig. 2b). These $V_{oc}$ values are far beyond its optical bandgap, indicating the presence of remarkable APV in **1**. Fig. S6 show the variation of $V_{oc}$ and $J_{sc}$ values measured in the intensity range from 58 to 149 $mW/cm^2$. It is evident that both $V_{oc}$ and $J_{sc}$ increase gradually with the enhancement of incident intensity and eventually reach saturation. To evaluate the reproducibility of the APV behavior, we conducted statistical measurements of $V_{oc}$ under identical illumination conditions. For eight different devices with channel lengths ranging from 0.8 to 2.0 mm, the $V_{oc}$ values are distributed between 14 and 40 V, showing a distance-dependent trend, as illustrated in Fig. 2c. Moreover, temperature-dependent $V_{oc}$ and $J_{sc}$ signals of device with channel length of ~1.0 mm were observed below 338 K ($T_c$), which is consistent with the temperature dependence of $P_s$ (Fig. 2d). This result confirms that the APV response in **1** is closely related to the internal electric field induced by ferroelectric polarization.

**Polarization-sensitive dependence of APV effect**. We further investigated the polarization-sensitive dependence of this ferroelectric APV effect in **1** under the polarized-light illumination. As shown in Fig. 3a, the incident beam sequentially passed through a polarizer and a half-wave plate (HWP) before illuminating onto the (0 1 0) crystal plane. Keeping the incident power constant, the polarization angle of incident light was altered by rotating the HWP. As shown in Fig. 3b, under a zero bias voltage and light intensity ranging from 58 to 149 mW/cm², the angular resolution $J_{sc}$ of ferroelectric APV exhibits a periodic variation of 180°. According to the bulk photovoltaic effect theory and Fridkin third-order tensor model, $J_c = I\beta_{31}\sin^2(\theta) + I\beta_{33}\cos^2(\theta)$ [18], the $J_{sc}$ exhibits strong polarization-angle dependence, reaching its minimum value ($J_{min}$) when the linear polarization aligns with the *a*-axis (0° polarization, $P_{s\perp}$). Maximum photocurrent ($J_{max}$) occurs at *c*-axis alignment (90° polarization, $P_{s\parallel}$), as shown in Fig. 3c. The polarization-sensitive dependence on light intensity indicates that the polarization ratio increases with the enhancement of incident power, and reaches its maximum at 128 mW/cm². As a result, **1** shows a large current ratio of $J_{max}/J_{min} \approx 41$, which far exceeds those of other 2D single-phase materials, such as black phosphorus (~8.7), (3-bromopropylammonium)$_2$PbBr$_4$ (~6.8), and (4-aminomethyl-1-cyclohexanecarboxylate)$_2$CsPb$_2$Br$_7$ (~2.0, Fig. 3d and Tab. S1) [19]. Consequently, the large polarization-sensitive ratio observed in **1** may relate to the coupling between the polarized-light and its above-bandgap photovoltage. The work reveals the enormous potential of 2D perovskite ferroelectric APV materials in high-performance polarized-sensitive optoelectronic applications.

**Shift current mechanism of ferroelectric APV.** To shed light on the origin of polarized-light response induced by ferroelectric APV effect in **1**, we employed DFT calculations to obtain the electronic structure and optoelectronic properties (cf. Methods see SI) [20, 21], as shown in Fig. 4. From the band-resolved contributions, the dominant terms originate from transitions between the second-highest valence band (VB$_2$) and the fourth conduction band (CB$_4$) (Fig. 4a). Fig. 4b illustrates the partial density of states (PDOS) and the charge density distributions for the CB$_4$ and VB$_2$ states at the Γ point. The VB$_2$ bands are derived from the Br *p*-orbitals, while the CB$_4$ band edges are largely from the Bi *p*-orbitals. When the electrons are excited from the Br-centered VB$_2$ states to the Bi-centered CB$_4$ states, a shift in the real space is expected to create the shift current. Figs. 4c and S7 show the shift current with respect to the photon energy along the crystallographic *c*-axis under linearly polarized light with its electric field vector along the *b*-axis, together with two dominant contributions based on the CB/VB combinations. The total shift current increases with the rising photon energy and reaches a maximum value of −5.8 μA/V² at 2.45 eV, which is comparable to that in BaTiO$_3$ (about 5 μA/V²) [22]. This significant peak near the bandgap energy highlights the strong nonlinear optical response in **1** and its anisotropy. That is, the shift current response observed along the *c*-axis direction is significantly stronger than that along the *a*- and *b*- axes directions, which is consistent with the large polarization ratio measured experimentally (Fig. S8).

On the other hand, it is well known that the shift current is of the topological nature [23], arising from the net displacement of excited electrons as the excited states evolve under an asymmetric crystal potential. The frequency-dependent response tensor ($\sigma$) can be expressed through shift vectors and transition dipole moments [24]:

$$\sigma_a{}^{bc}(\omega) = e \sum_{n,m} \int d\mathbf{k}\, I^{bc}(m,n,\mathbf{k};\omega) R_a(m,n,\mathbf{k}) \qquad (4)$$

$$I^{bc}(m,n,\mathbf{k};\omega) = \pi \left(\frac{e}{m\hbar\omega}\right)^2 f_{nm}(\mathbf{k}) \mathbf{P}_{mn}^b \mathbf{P}_{nm}^c \times \delta[\omega_{nm}(\mathbf{k}) \pm \omega] \qquad (5)$$

$$R_a(m,n,\mathbf{k}) = \partial_{k_a}\phi_{nm}(\mathbf{k}) - [A_{na}(\mathbf{k}) - A_{ma}(\mathbf{k})] \qquad (6)$$

where $I$ represents the transition intensity, and $R$ denotes the shift vector, which characterizes the instantaneous carrier displacement during the optical excitation. $A$ denotes the Berryconnection, and $\phi$ represents the transition dipole moment's quantum phase. The terms $\omega_{nm}$ and $f_{mn}$ correspond to the energy difference and occupation difference between the involved bands, respectively.

Consequently, the shift current emerges as an integral over each $k$-point in the Brillouin zone. At each $k$-point, the contribution is given by the product of the transition intensity and the shift vector summed over the relevant band pairs. The $k$-resolved shift current distribution at $\hbar\omega = 2.45$ eV is displayed in Fig. 4d. It is evident that, in the momentum space, the major contribution originates from the region around the $\Gamma$ point and along the high-symmetry $\Gamma{\to}S$ line. Moreover, the shift current exhibits a distribution pattern similar to that of the shift vector. In regions with significant transition intensity, the peaks of the $k$-resolved shift current coincide with those in the shift vector. In these regions, all shift vectors share the same sign, ensuring their contributions can not cancel out. Such displacements effectively enhance the overlap between the shift vector and transition intensity across the Brillouin zone, leading to a pronounced amplification of the shift current response [25]. Notably, the shift current enables the open-circuit voltage to surpass the bandgap limitation through the coherent superposition of quantum geometric displacements and Berry phase modulation, independent of thermal equilibrium, which coincides fairly well with the APV performances in **1**.

## Discussion

Subsequently, we performed a detailed analysis of the ionic displacements within the structure of **1**. As shown in Fig. 5a, the Cs$^+$ cations in the crystal lattice of **1** deviate from the center of perovskite cage with an average displacement ~0.144 Å along the $c$-axis direction. Therefore, compared with the 3D prototype Cs$_2$AgBiBr$_6$, Cs$^+$ cations exhibit a lower migration energy barrier in **1** and serve as the dominant ions with dynamic feature [26]. In particular, light illumination facilitates the expansion of the lattice of perovskite framework, which might weaken the electrostatic interaction between Cs$^+$ cations and AgBr$_6$/BiBr$_6$ octahedra [27]. This effectively reduces the energy barrier of ion migration and promotes ion migration of $Cs^+$ cations (Fig. 5b). To confirm the ion migration behavior of **1**, we measured the photovoltage *versus* time (*V-t*) curves of device with channel length of ~1.0 mm under different pre-illumination. As shown in Figs. 5c and 5d, in the initial state without pre-illumination, the $V_{oc}$ is about 20 V. Then, the same power of illumination was applied to the same device for 4-20 min to ensure the generation of ion migration. With the increasing of pre-illumination time, the $V_{oc}$ decreases slightly, and the $J_{sc}$ gradually decreases from 14 to 12 $nA/cm^2$. In addition, the response time of $V_{oc}$ and $J_{sc}$ also significantly increases after pre-illumination. Particularly, the amplification of rise ($\tau_{rise}$) and decay ($\tau_{decay}$) times are about 460% and 250%, respectively. It demonstrates that the ion migration process forms more vacancy defects, which trap more photocarriers [28]. The results show that $Cs^+$ cations displacement plays an important role in inducing ferroelectric polarization and the shift current caused by ion migration, thereby providing a crucial driving mechanism for the APV effect.

**Conclusion**

In summary, we have reported strong APV activity in a lead-free double perovskite ferroelectric $(CHMA)_2CsAgBiBr_7$, which shows a large above-bandgap photovoltage up to ~40 V, much higher than its band gap. Interestingly, the shift current model closely associated with $Cs^+$ cation displacement has for the first time unveiled the structural origin of ferroelectric APV activity. Further, the steady-state APV photocurrents of **1** exhibit remarkable polarization-sensitivity with a record-high polarization ratio of ~41 among known 2D single-phase materials. This work enriches the ferroelectric APV candidates and provides inspire for the design of high-performance optoelectronic devices.

**Acknowledgments**

This work was supported by the National Natural Science Foundation of China (22125110, 22205233, U23A2094, 22305248 and 22405271), the Natural Science Foundation of Fujian Province (2023J02028), the Key Research Program of Frontier Sciences of the Chinese Academy of Sciences (ZDBS-LY-SLH024), the Strategic Priority Research Program of the CAS (XDB20010200). The Lichtenberg high performance computer of the TU Darmstadt is gratefully acknowledged for the computational resources where the calculations were conducted for this project.

**Materials and Methods**

**Synthesis of crystal**. Cyclohexylmethanamine, cesium carbonate, $Ag_2O$ and $Bi_2O_3$ were mixed in concentrated hydrobromic acid solution according to stoichiometric ratio, and yellow crystal compound **1** was prepared by temperature cooling method (Fig. S1).

**Electrical measurements**. The temperature-dependent *P-E* hysteresis loops and *J-E* current curves were carried out on a ferroelectric analyzer (Radiant Precision Premier II).

**Optical measurements**. The UV absorption spectrum in the solid state was measured at room temperature on a PE Lambda 900 UV-visible-NIR spectrophotometer. A Keithley 6517B source meter and a series of filters for wavelength-dependent measurement were adopted on a lateral structure device.

**Computational methods**. The DFT calculations were performed using the projected augmented plane-wave method as implemented in the Vienna Ab initio Simulation Package (VASP). The exchange-correlation functional is treated within the generalized gradient approximation of Perdew-Burke-Ernzerhof (PBE) type.

**Figures**

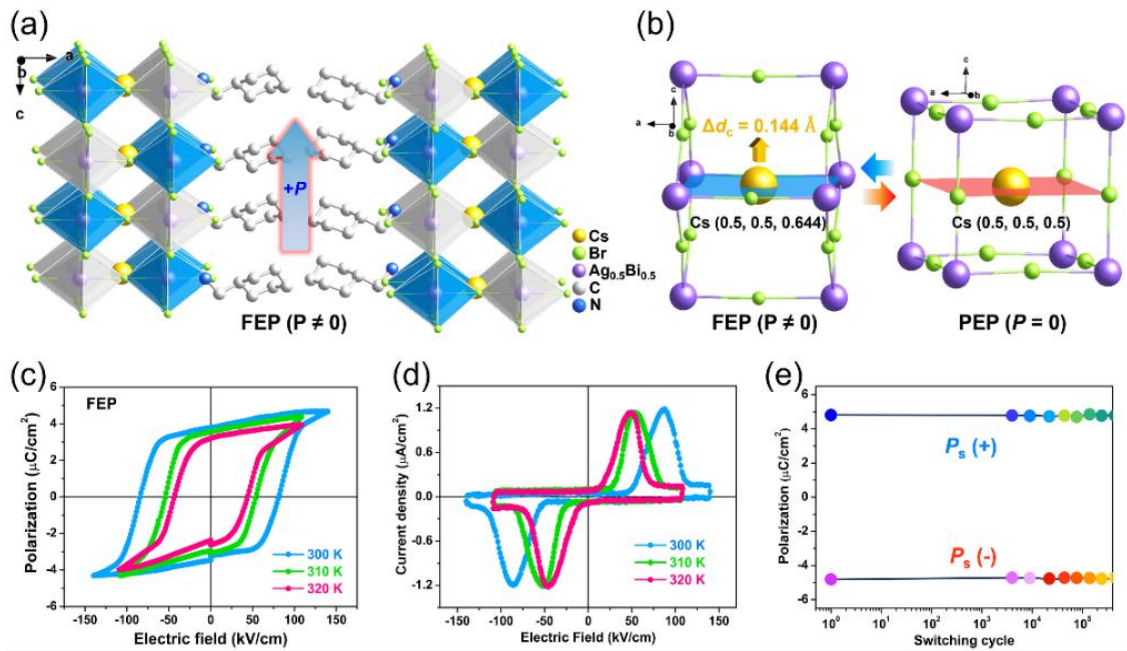

**Fig. 1**. (a) Packing diagrams of **1** in the ferroelectric phase (FEP). (b) Schematic diagram for atomic displacements of Cs⁺ cations along the *c*-axis direction. (c) *P-E* hysteresis loops and (d) *J-E* curves at different temperatures. (e) The fatigue characteristic of **1** after ~4 × 10⁵ switching cycles at 300 K.

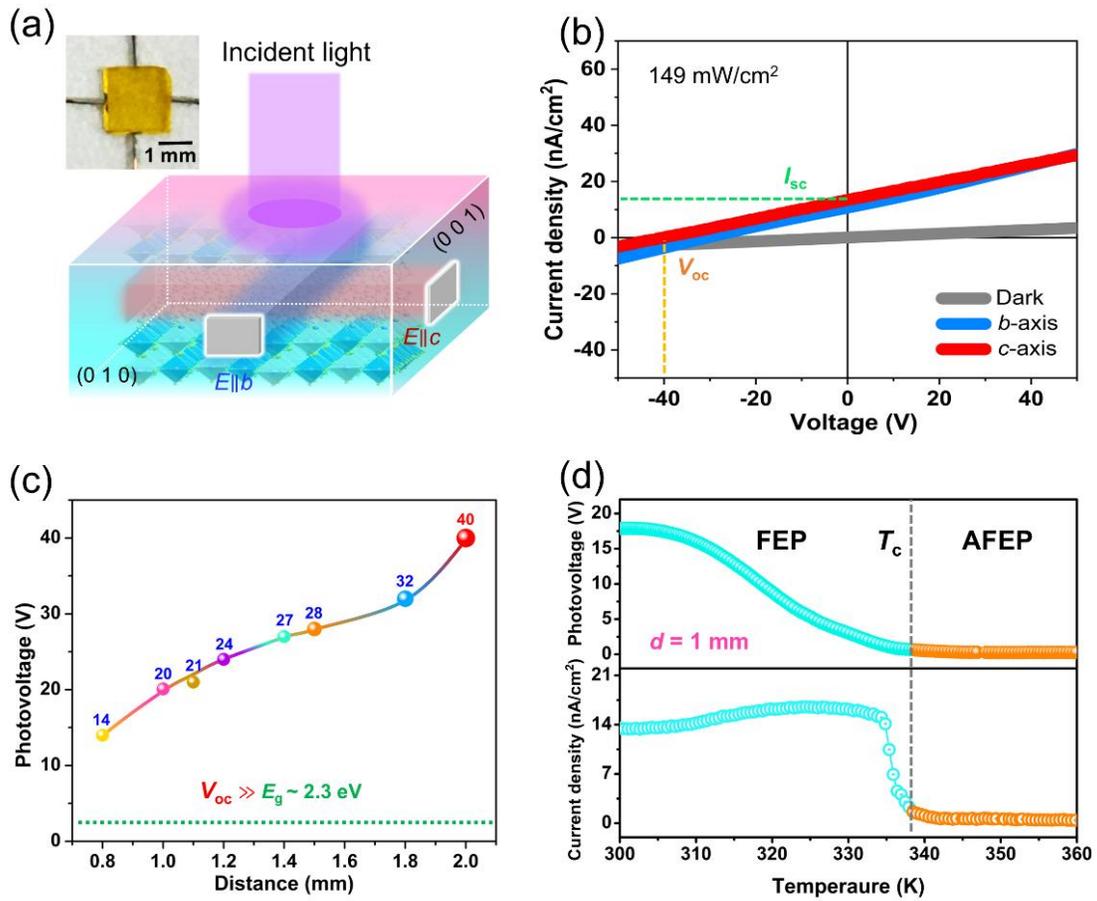

**Fig. 2**. (a) Schematic of bulk photovoltaic effect measurements and optical image of the sample. (b) $J$-$V$ curves measured along the directions parallel to the $c$- and $b$-axes, respectively. (c) Statistics of the measured $V_{oc}$ for eight devices with spacings from 0.8 to 2.0 mm. (d) Temperature-dependent $V_{oc}$ and $J_{sc}$ of **1**.

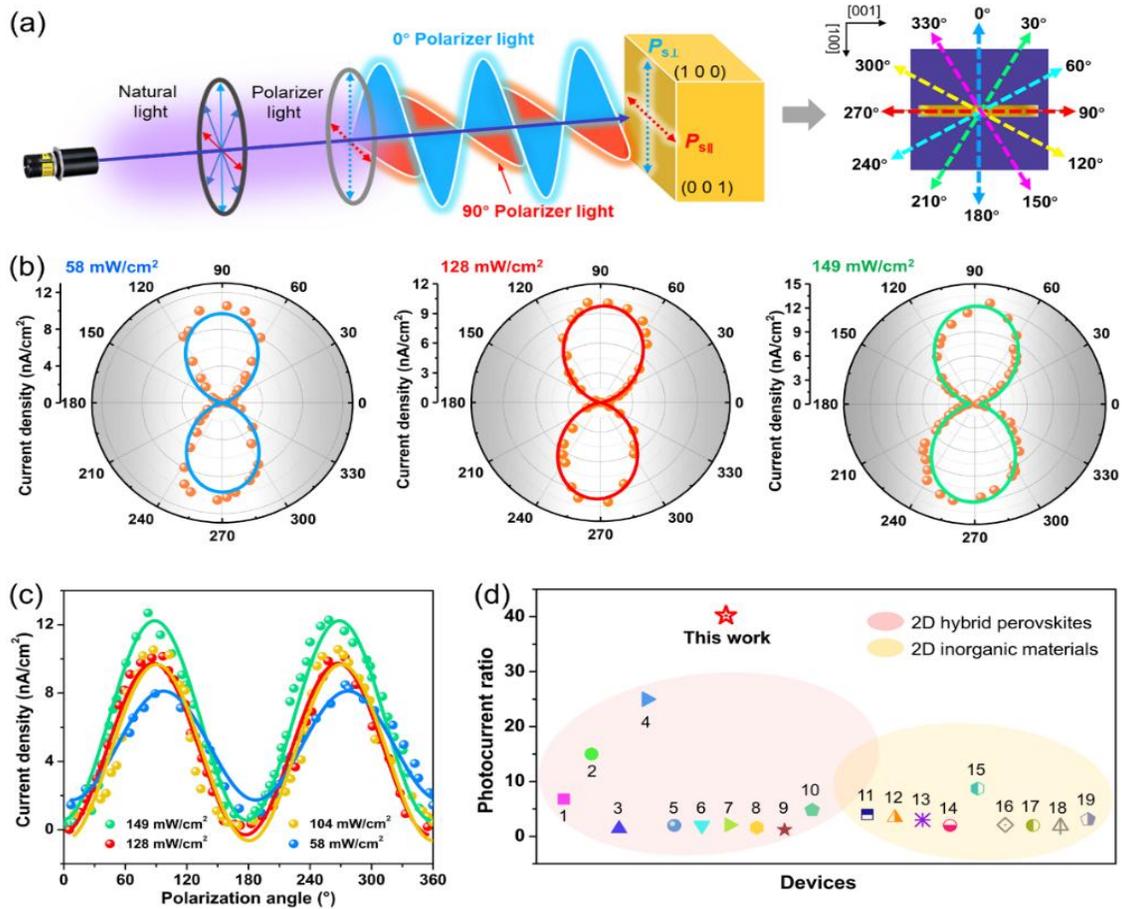

**Fig. 3**. (a) Schematic of polarized-light detector. (b) Polarization dependence of $I_{sc}$ and (c) angle-resolved $I_{sc}$ collected under different illumination intensities. (d) The polarization ratio for various reported 2D phototronic detectors (1: $(BPA)_2PbBr_4$; 2: $(AA)_2(EA)_2Pb_3Br_{10}$; 3: ($i$-PA)$_2$CsAgBiBr$_7$; 4: $(HA)_2(EA)_2Pb_3Br_{10}$; 5: $(TRA)_2CsPb_2Br_7$; 6: $(BA)_2(MA)Pb_2Br_7$; 7: $(FPEA)_2PbI$; 8: ($i$-BA)PbI$_4$; 9: ($i$-BA)$_2$(MA)Pb$_2$I$_7$; 10: $(BDA)(EA)_2Pb_3Br_{10}$; 11: ReS$_2$; 12: $\beta$-GeSe; 13: GeSe; 14: GeAs$_2$; 15: Black phosphorus; 16: PdSe$_2$; 17: GeS$_2$; 18: Td-TaIrTe$_4$; 19:$o$-SiP).

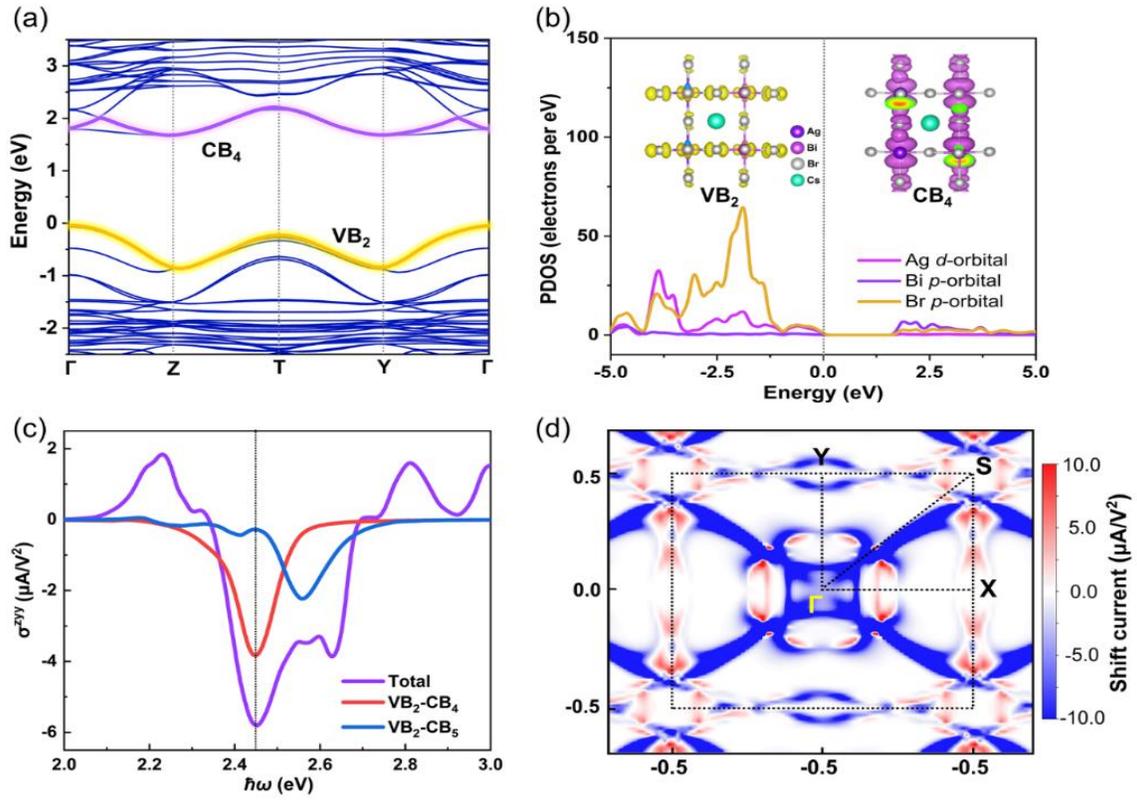

**Fig. 4**. (a) Band structure of **1**. (b) The total shift current and band-resolved dominant contributions. The calculated shift current spectra σ along the x, y, and z directions, corresponding to the crystallographic *a*-, *b*-, and *c*-axes, respectively. (c) Partial density of states (PDOS), with insets showing the charge density distributions for the CB₄ and VB₂ of **1** at the Γ point. (d) The *k*-resolved shift current at $\hbar\omega$ = 2.45 eV.

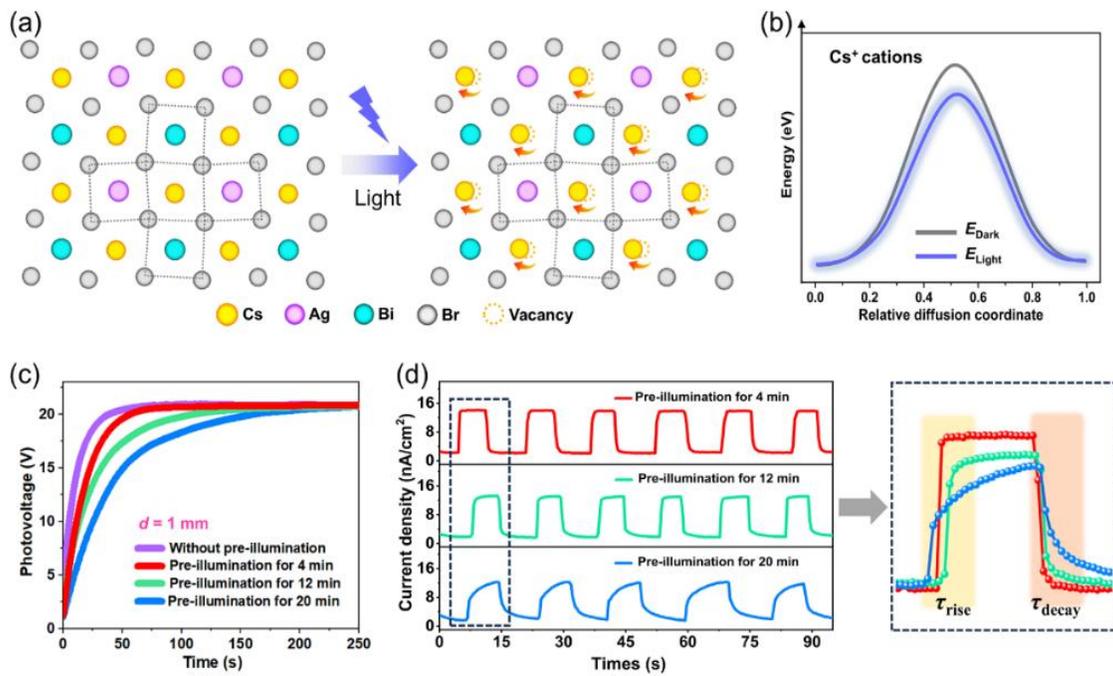

**Fig. 5**. Schematic diagram for (a) atomic displacements and (b) energy profiles of ion migrations of Cs$^+$ cations. (c) $V$-$t$ and (d) $J$-$t$ curves under different pre-illumination durations.

# Table of Contents

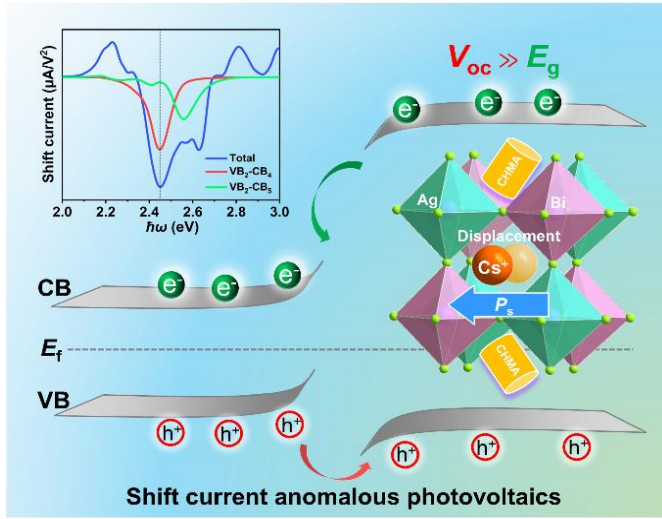